\documentclass[amssymb,pra,twocolumn,notitlepage,longbibliography,superscriptaddress]{revtex4-1}

\renewcommand{\vec}[1]{\mathbf{#1}}

\newcommand{\kets}[1]{| #1 \rangle}
\newcommand{\bras}[1]{\langle #1 |}

\usepackage[%
colorlinks=true,
urlcolor=blue,
linkcolor=blue,
citecolor=blue
]{hyperref}

\usepackage{color}

\usepackage{amsmath}                          % better math control.
\usepackage{amsfonts}                         % extended math fonts.
\usepackage{amssymb}                          % names amsfont symbols.
\usepackage{amsthm}                           % theorem environments
\usepackage{mathdots}                         % for ddots going upwards

\usepackage[braket, qm]{qcircuit}   % for drawing quantum circuit

\usepackage{graphicx}                         % ps, pdf figs.
\usepackage{subfigure}                        % for subfigures
\usepackage{overpic}

\begin{document}
%\title{Variational quantum eigensolver with cooperative neural networks}
%\title{Cooperative variational quantum eigensolvers}
\title{Collective optimization for variational quantum eigensolvers}
\author{Dan-Bo Zhang}
\email{dbzhang@m.scnu.edu.cn}
\affiliation{Guangdong Provincial Key Laboratory of Quantum Engineering and Quantum Materials, GPETR Center for Quantum Precision Measurement, SPTE, South China Normal University, Guangzhou 510006, China}
\author{Tao Yin}
\email{tao.yin@artiste-qb.net}
\affiliation{Yuntao Quantum Technologies, Shenzhen, 518000, China}
%\affiliation{YunTao Quantum Tech}
%\affiliation{Guangdong Provincial Key Laboratory of Quantum Engineering and Quantum Materials, GPETR Center for Quantum Precision Measurement, SPTE, South China Normal University, Guangzhou 510006, China}
%\author{Shi-Liang Zhu} \email{slzhu@nju.edu.cn}
%\affiliation{Guangdong Provincial Key Laboratory of Quantum Engineering and Quantum Materials, GPETR Center for Quantum Precision Measurement,  SPTE, South China Normal University, Guangzhou 510006, China}
%\affiliation{National Laboratory of Solid State Microstructures, School of Physics, Nanjing University, Nanjing 210093, China}
%
%\author{Z. D. Wang}simultaneously
%\email{zwang@hkucc.hku.hk}
%\affiliation{Department of Physics and Center of Theoretical and Computational Physics, The University of Hong Kong, Pokfulam Road, Hong Kong, China}
%\affiliation{Guangdong Provincial Key Laboratory of Quantum Engine ering and Quantum Materials, GPETR Center for Quantum Precision Measurement,SPTE, South China Normal University, Guangzhou 510006, China}

\begin{abstract}
%Variational quantum eigensolvers (VQE) aim to solve quantum systems by preparing eigenstates with parameterized quantum circuits. The parameters are obtained in the framework of hybrid quantum-classical computation by optimizing system energies, and thus one centric issue is to find effective optimization algorithms, such that can avoid local minimums or save quantum resources. Here, we incorporate a snake algorithm to optimize a series of related VQE tasks collectively, which trends to pull parameters of single task out of \textcolor{red}{traps of} local minimums. Then such so called collective VQE~(cVQE) is applied for solving molecules with various bond distances simultaneously. The collective optimization exploits intrinsic relations between realted tasks and may inspire  advanced hybrid quantum-classical algorithm for solving practical problems on near-term quantum devices.
Variational quantum eigensolver (VQE) optimizes parameterized eigenstates of a Hamiltonian on a quantum processor by updating parameters with a classical computer. Such a hybrid quantum-classical optimization serves as a practical way to leverage up classical algorithms to exploit the power of near-term quantum computing. Here, we develop a hybrid algorithm for VQE, emphasizing the classical side, that can solve a group of related Hamiltonians simultaneously. The algorithm incorporates a snake algorithm into many VQE tasks to collectively optimize variational parameters of different Hamiltonians. Such so-called collective VQEs~(cVQEs) is applied for solving molecules with varied bond length, which is a standard problem in quantum chemistry. Numeral simulations show that cVQE is not only more efficient in convergence, but also trends to avoid single VQE task to be trapped in local minimums. The collective optimization utilizes intrinsic relations between related tasks and may inspire advanced hybrid quantum-classical algorithms for solving practical problems.
\end{abstract}
\maketitle

\section{Introduction}
%Why we should care about VQE
%why we should consider a group of Hamiltonians
%review of snake algorithm and argue why it can use for collective optimization
%claim what we have done
Quantum computing exploits intrinsic quantum properties for computing. It promises to solve some outstanding problems with quantum advantages\cite{Feynman_1982,shor_97,harrow_09,aaaronson_14,bravyi_18,Arute2019}, and is influencing a broad of computational intensive areas, such as quantum simulation~\cite{Feynman_1982,abrams_97,buluta_09,trabesinger_12} and machine learning~\cite{biamonte_17}. A variational approach for quantum computing sets parameters in a quantum circuit, and learn those parameters through hybrid quantum-classical optimization~\cite{yung_14,farhi_14,mcclean_16,o’malley_16,li_17,mcclean_17,shen_17,kandala_17,hempel_18,anschuetz_18,mitarai_18,moll_18,kokail_19,takeshita_19,mcardle_19,Higgott_19,sweke_19}. Such an approach is well suited for near-term quantum processor, and receives lots of attention in recent years~\cite{Preskill_18}. Among them, variational quantum eigensolver (VQE) aims to solve eigenvalues and eigenstates for quantum systems~\cite{yung_14,mcclean_16,kandala_17,hempel_18,liu_19}. The power of representing exponentially large wavefunction on quantum processors and effective hybrid quantum-classical optimization of VQE enhances the ability to solve hard quantum problems.

In many practical problems, a group of related Hamiltonians needs to be solved. For instance, molecule electronic Hamiltonians under different bond lengths or angles, or quantum many-body systems with different interacting strengths. VQE can solve such a group of Hamiltonians one by one independently, without taking advantage of previous results. However, those tasks are mostly similar and related to each other, such that one can exploit intrinsic relations for more efficient optimization that can require less quantum resources or avoid local minimums for single tasks. This is also related to meta learning that draws prior experience for new tasks~\cite{andrychowicz_16,RN1258,verdon_19}.
% however, "reduce computational resource in optimization" is overclaimed. should be rephased.

In this paper, we propose a hybrid quantum-classical algorithm that can provide a collective optimization for VQE to solve a group of related Hamiltonians simultaneously. It evaluates gradients on the quantum processor and updates variational parameters on the classical computer. Remarkably, the updating process generalizes typically gradient descent into a collective version, which updates variational parameters of different Hamiltonians simultaneously. This is achieved by a snake algorithm~\cite{kass_88,liu_18}, originally developed in computer vision~\cite{kass_88}, which enforces a smooth condition on variational parameters of different Hamiltonians. We call this collective VQE or cVQE. As demonstrations, we use the cVQE to solve ground-state energies for several molecules at different bond lengths. The advantages of collective optimization are investigated and shown through the flow of variational parameters. Remarkably, the snake algorithm is revealed as a global optimizer, as collective motion of parameters for different tasks can pull a point of parameters for a single Hamiltonian out of traps of local minimums.

The paper is organized as follows. In Sec.~\ref{sec:review}, we review variational quantum eigensolver, and then propose cVQE using the snake algorithm. In Sec.~\ref{sec:application}, we present results of several representative molecules using cVQE. In Sec.~\ref{sec:global_optimizer}, we investigate the snake algorithm as a global optimizer. Finally, we give some further discussions and a brief summary.

\section{Optimization for variational quantum eigensolvers} \label{sec:review}
Solving eigenvalues and eigenstates for a given Hamiltonian is a basic task. %especially on ground state and ground state energies. %This is also of practical interest for quantum chemistry, where chemical properties largely depend on the ground state with a gap that is much larger than $25.7$eV (representative energy scale at room temperature).
%Solving quantum chemistry or quantum many-body problems is considered to be QMA-hard~\cite{aaronson_09}, which is believed to be even harder than NP-hard problem.
Quantum computers provide an avenue for solving eigenstate problems of quantum systems effectively. Different quantum algorithms have been developed for tracking this hard problem, such as quantum phase estimation~\cite{aspuru-guzik_05}, variational quantum eigensolver~\cite{yung_14,mcclean_16}, simulating resonance transition of molecules on quantum processors~\cite{wang_12,li_19}. The VQE approach uses a parameterized quantum circuit to prepare a wavefunction. The parameters are obtained by optimizing the energy with the hybrid quantum-classical algorithm.

To solve quantum systems on a quantum computer, it is necessary to firstly map the original Hamiltonian in a qubit~(spin-half) Hamiltonian.
For electronic systems, a nonlocal transformation such as Jordan-Wigner transformation~\cite{jordan_28} or Bravyi-Kitaev transformation~\cite{bravyi_02}, is required to firstly transform fermionic operators into Pauli operators. For a quantum system of interest (e.g., molecules), the resulting qubit Hamiltonian typically has many terms,
\begin{equation}
H=\sum_i c_i H_i,
\end{equation}
where $H_i$ can be written as a tensor product of Pauli matrices, $H_i= \otimes_k\sigma^{\alpha_k}_{k}$. Here $\alpha_k=x,y,z$ and $k$ the index of qubits. We now discuss how to solve a single Hamiltonian and a group of related Hamiltonians, respectively.

\subsection{Optimization by gradient descent}
To find the eigenstate for a single $H$, one can use an ansatz $\kets{\psi(\boldsymbol{\theta})}=U(\boldsymbol{\theta})\kets{\psi_0}$ to represent a candidate ground state. Here $\kets{\psi_0}$ is an initial state as a good classical approximation as the ground state of $H$. For instance, $\kets{\psi_0}$ can be chosen as a Hartee-Fock state in quantum chemistry.  $U(\boldsymbol{\theta})$ is an unitary operator parameterized with $\boldsymbol{\theta}$, which can take quantum correlation into consideration.
%The unitary $U(\boldsymbol{\theta})$ takesquantum correlations into consideration and thus provides a better approximation. It is parameterized with $\boldsymbol{\theta})$, and can be constructed with a parametrized quantum circuit, consisting of one-qubit and two-qubit gates. Parameters can appear only as as rotational angles of one-qubit gates. $U(\boldsymbol{\theta})$ can also be constructed from Hamiltonian evolutions with alternating Hamiltonians, and parameters can appear as a series of evolution interval times~\cite{farhi_14,kokail_19}. In this paper we mainly adopt the form approach .  Finding a proper ansatz $U(\boldsymbol{\theta})$ is a central issue for variational quantum eigensolver: the ansatz should have enough expressive power to include the true ground state in the subspace.
%on the other hand: parameters of global minimum can be found with optimization methods requiring affordable quantum resources.
As a variational method, the essential task is to find parameters $\boldsymbol{\theta}_0$ that minimizes the energy $\mathcal{E}(\boldsymbol{\theta})=\bras{\psi(\boldsymbol{\theta})}H\kets{\psi(\boldsymbol{\theta})}$.
The optimization is a hybrid quantum-classical one: the quantum processor runs the quantum circuit and performs measurements to evaluate $\mathcal{E}(\boldsymbol{\theta})$; the classical computer updates parameters $\boldsymbol{\theta}$ according to received data from the quantum processor. To obtain a quantum average of $H$, one can perform measurements for each term $H_i$, as it is a tensor product of Pauli matrices thus corresponds to a joint measurement on multi-qubits. Measurements of all terms then are added,
\begin{equation}
\mathcal{E}(\boldsymbol{\theta})=\sum_ic_i\bras{\psi(\boldsymbol{\theta})}H_i\kets{\psi(\boldsymbol{\theta})}.
\end{equation}
Optimization methods for updating parameters $\boldsymbol{\theta}$ in general can be  categorized as gradient free~\cite{hempel_18,kokail_19}, such as Nelder-Mead method, and gradient descent~\cite{kandala_17,liu_19,sweke_19}.
% The Nelder-Mead method is gradient free as it does not require information of gradient, and it has been adopted in VQE for simple quantum chemistry problems. Many other gradient free methods have been also developed.
Gradient descent methods update parameters using information of gradients. On a quantum processor, calculating gradient with respect to a target cost function (here is $\mathcal{E}(\boldsymbol{\theta})$) can be obtained with the same quantum circuit, using the shift rule~\cite{lijun_17,schuld_19} or numeral differential.
%assuming parameters appear only in single-qubit gates. \textcolor{red}{Explicitly , a gradient $\frac{\partial}{\partial{\boldsymbol{\theta}_i}}\mathcal{E}(\boldsymbol{\theta})$ ($\boldsymbol{\theta}_i$ is the $i$-th component of $\boldsymbol{\theta}$) has an analytical expression}
%\begin{equation}\label{eq:shift_rule}
%\frac{\partial}{\partial{\boldsymbol{\theta}_i}}\mathcal{E}(\boldsymbol{\theta})=\frac{1}{2}(\mathcal{E}(\boldsymbol{\theta}+\frac{\pi}{2}\boldsymbol{e}_i)-\mathcal{E}(\boldsymbol{\theta}-\frac{\pi}{2}\boldsymbol{e}_i)),
%\end{equation}
%\textcolor{red}{where $\boldsymbol{e}_i$ is an unit vector of the $i$-th component.}
Then parameters $\boldsymbol{\theta}$ are updated with gradient descent as
\begin{equation}\label{eq:gd}
\boldsymbol{\theta}^{t}=\boldsymbol{\theta}^{t-1}-\eta \frac{\partial}{\partial{\boldsymbol{\theta}}}\mathcal{E}(\boldsymbol{\theta}^{t-1}),
\end{equation}
where $\eta$ is the learning rate or step size.

%It is should be pointed out that a simultaneous perturbation stochastic approximation (SPSA) usually is adopted for optimization on a noisy quantum processor, where gradients are taken for randomly chosen directions~\cite{kandala_17}.

\subsection{Collective optimization}
\label{sec:snake_algorithm}
In the above, variational quantum eigensolver solves the eigenvalue problem for a single Hamiltonian. In practice, there may be a group of Hamiltonians to be solved. For instance, what is needed in quantum chemistry usually is a potential surface, corresponding to ground state energies for a molecule at different bond lengths or bond angles.
%In quantum many-body systems, energies of the ground state  and the first excited state  should be calculated with varying parameters in the Hamiltonian to detect if there is a quantum phase transition.
Of course, one can  use VQE to solve Hamiltonians one by one. However, this does not exploit relations between Hamiltonians. Here, we develop a more efficient method that can collectively optimize
all variational gate parameters for different Hamiltonians at the same time.

The motivation behind collective optimization can be presented as follows. Consider quantum chemistry problems. Two Hamiltonians should be close to each other if their underlying molecules are the same and bond lengths vary a little. In such a case, the same ansatz can be applied, and it is expected that optimized parameters of wavefunction should be very close to each other. Denoted $\boldsymbol{\theta}_0(\lambda)$ as the optimized parameter for Hamiltonian $H(\lambda)$, then $\boldsymbol{\theta}_0(\lambda) \sim \lambda $ should form a continuous curve in the space of $\boldsymbol{\theta}$ and $\lambda$, which we call as enlarged parameter space. We expect that the optimization of one Hamiltonian can help optimize other Hamiltonians with nearby system parameters $\lambda$.
We use gradient descent for the optimization. Instead of updating a single point in the parameter space, the optimization updates a sequence of points in the enlarged parameter space, each point corresponding to a Hamiltonian. At the continuous limit, this is an optimization of a string.
%There are several benefits. Firstly, it can speed up the process of calculating potential surface; secondly, it may overcome vanishing gradients problem that may be met when optimizing variational quantum circuits, as collective optimization can provide better initializations.
%lastly, collective optimization may be more robust to statistical error (thus require less measurements per Hamiltonian).

%\subsubsection{The snake algorithm}
Now let us elaborate on a concrete algorithm.
%, which has been developed in computer vision for edge or contour detection. Here, the task is to optimize a string $\boldsymbol{\theta}$ in the enlarged parameter space $(\lambda,\boldsymbol{\theta})$, where the string can be considered as an edge. The cost function~(action)
To incorporate a snake algorithm, the cost function should consider energy of the snake itself, and can be written as follows:
\begin{equation}\label{eq:action}
L[\boldsymbol{\theta}(\lambda)]=\int_{\lambda_0}^{\lambda_T}(\mathcal{L}(\boldsymbol{\theta}(\lambda))+\mathcal{E}(\boldsymbol{\theta}(\lambda)))
\end{equation}
Here $\mathcal{E}(\boldsymbol{\theta}(\lambda))=\bras{\psi(\boldsymbol{\theta}(\lambda))}H_\lambda\kets{\psi(\boldsymbol{\theta}(\lambda))}$ is the local potential the snake feels and the internal property is
\begin{equation}
\mathcal{L}(\boldsymbol{\theta}(\lambda))=\alpha |\frac{\partial\boldsymbol{\theta}(\lambda)}{\partial\lambda}|^2+\beta |\frac{\partial^2\boldsymbol{\theta}(\lambda)}{\partial^2\lambda}|^2,
\end{equation}
where $\alpha$ and $\beta$ terms make the snake stretchable and bendable~\cite{kass_88,liu_18}, respectively.

Solving the snake can be achieved by minimizing Eq.~\ref{eq:action}, which can converted to solve a differential equation(see Eq.~\ref{eq:snake-diff} in the Appendix). For this we discrete the snake as a sequence of parameters at different bond lengths
$\vec{r}_i=(\boldsymbol{\theta}_i(\lambda_1),\boldsymbol{\theta}_i(\lambda_2),...,\boldsymbol{\theta}_i(\lambda_M))$, where $i$ is the $i$-th component for each $\boldsymbol{\theta}_i(\lambda_m)$. Then, the discrete snake can be solved iteratively as
\begin{equation}\label{eq:collective_gd}
\vec{r}_i^{t}=\left( \eta\boldsymbol{A}+\boldsymbol{I} \right)^{-1}\left( \vec{r}_i^{t-1}-\eta\frac{dE(\vec{r}^{t-1})}{d\vec{r}_i} \right)
\end{equation}
where $E(\vec{r})=\sum_m\mathcal{E}(\boldsymbol{\theta}_{\lambda_m})$, and $\boldsymbol{A}$ is a pentadiagonal banded matrix with nonzero elements depending on $\alpha$ and $\beta$.
Details can be found in the Appendix~\ref{app:colllective_gd}.
Compared with Eq.~\eqref{eq:gd}, Eq.~\eqref{eq:collective_gd} can be viewed as a collective gradient descent, as the later is reduced to the former at $\alpha=\beta=0$.

There is an issue for incorporating the snake algorithm into optimizing variational quantum eigensolver. The equilibrium condition Eq.~\eqref{eq:matrix_pde}~(or Eq.~\eqref{eq:snake-diff}) is actually not the original one
$\frac{d\mathcal{E}(\boldsymbol{\theta}(\lambda_i))}{d\boldsymbol{\theta}(\lambda_i)}=0
$,
as there are interactions between neighbor $\boldsymbol{\theta}(\lambda_i)$. As a result, optimization with a gradient flow using Eq.~\eqref{eq:collective_gd} may not give the required optimal results. In practice, nevertheless, this issue may be largely ignored, as explained in the following. For neighbor $\lambda_i$,
it can expected that $\boldsymbol{\theta}(\lambda_{i+1})+\boldsymbol{\theta}(\lambda_{i-1})\approx2\boldsymbol{\theta}(\lambda_i))$ and $\boldsymbol{\theta}(\lambda_{i+2})+\boldsymbol{\theta}(\lambda_{i-2})\approx2\boldsymbol{\theta}(\lambda_i))$ once the optimization is good enough and $M$ is large enough. It can be checked that the first term of Eq.~\ref{eq:matrix_pde} can be approximated as zero, which is consist with the equilibrium condition for VQE, namely by omitting the first term.

In practice, we can introduce a decaying matrix $\boldsymbol{A}(t)=\boldsymbol{A}_0\exp(-t\Gamma)$ in the optimization process.
%The gradient flow  Eq.~\ref{eq:collective_gd} becomes
%\begin{equation}\label{eq:collective_gd_dynamical}
%\vec{r}_i^{t}=\left( \eta\boldsymbol{A}_0\exp(-t\Gamma)+\boldsymbol{I} \right)^{-1}\left( \vec{r}_i^{t-1}-\eta\frac{dE(\vec{r}^{t-1})}{d\vec{r}_i} \right).
%\end{equation}
For large $t$ limit, this become the gradient descent of Eq.~\ref{eq:gd}. An analog may be made with the annealing methods widely applied for optimization. Internal forces play the role of temperature. Initially, internal forces are large and parameters for different Hamiltonians flow in the space collectively. With decaying internal forces flows of different parameters become more independent. This may inspire us that the snake algorithm may help avoid the optimization to be trapped in a local minimum for a single VQE, which will be investigated at Sec.~\ref{sec:global_optimizer}.

%\subsubsection{Open boundary condition}
%Some illness corresponds to localized modes.

\section{Application of cVQE for molecules}
\label{sec:application}
In this section, we apply cVQE for several representative molecules, including molecular hydrogen, Lithium hydride and Helium hydride cation and present their results. The numerical simulations are performed by using Huawei HiQsimulator framework~\cite{HiQ}. It is shown that ground state energies are obtained with great accuracy compared with results using variational quantum eigensolver for Hamiltonian at each bond length alone. Remarkably, variational parameters for ground states of Hamiltonians at different bond lengths collectively flow to optimal values. We present the main results and details of the calculation of Hamiltonians for all molecules at different bond lengths as well as their wavefunction ansatz are put in Appendix.~\ref{appdix:ham_openfermion}.

\begin{figure*}
	\includegraphics[width=1\textwidth]{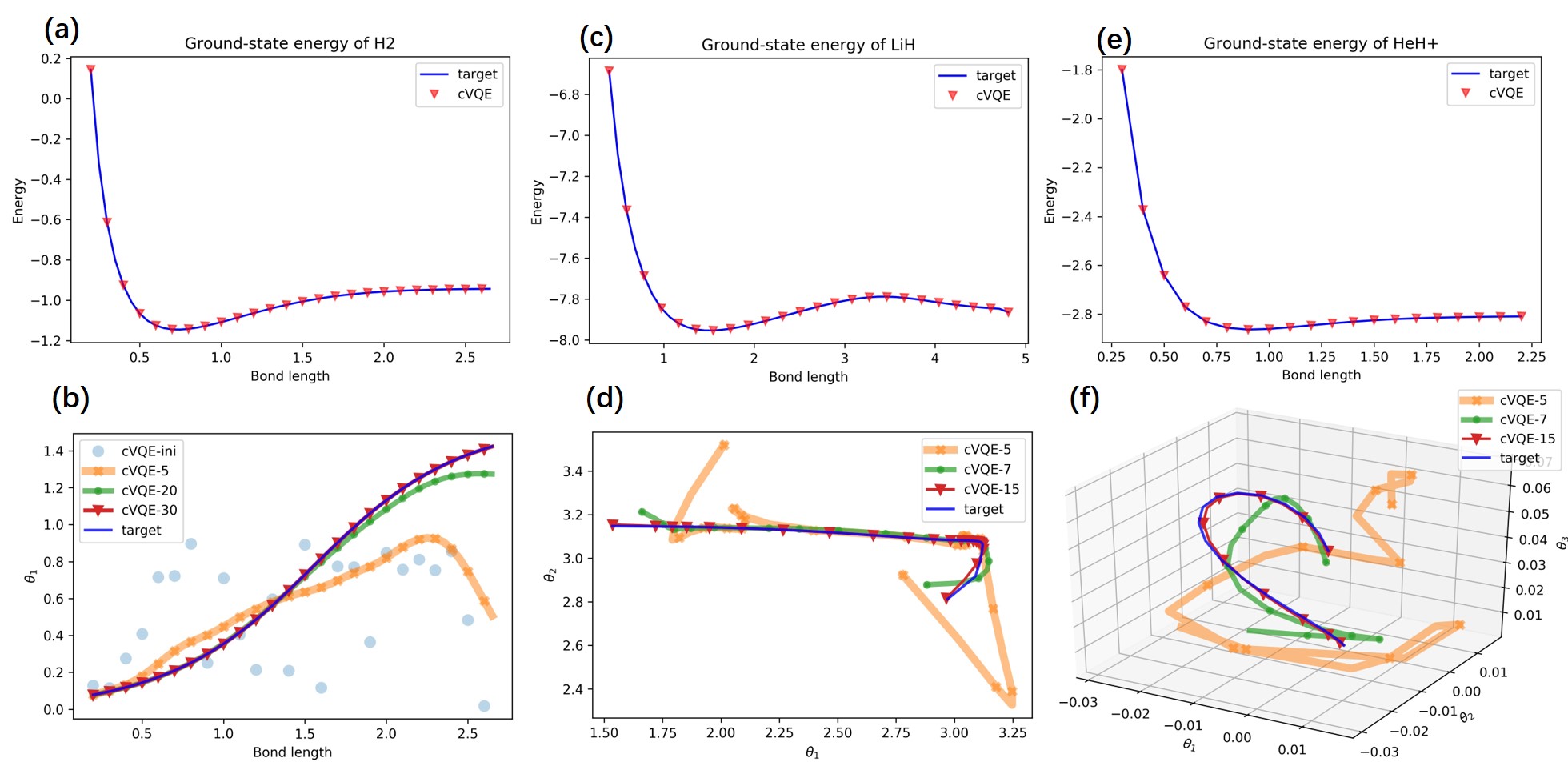}
	\caption{Collective optimization with the snake algorithm for molecules at different bond lengths. The first row shows the target energies and optimization results for the cVQE algorithm, and the second row presents optimization process of variational parameters. The left, the middle and the right columns correspond molecules $\text{H}_2$, $\text{LiH}$ and $\text{HeH}^+$, with one, two, and three variational parameters, respectively. }
	\label{fig:snake_algorithm}
\end{figure*}

\subsection{Molecular hydrogen}
%For molecular hydrogen, the UCC operator $U(\theta)=\exp(-i\theta\sigma_0^x\sigma_1^y)$
%%performs on an initial state $\kets{01}$.
%To apply $U(\theta)$ on a quantum processor, it is necessary to decompose it into a combination of standard universal quantum gates. With CNOT and single-qubit rotation gates, $U(\theta)$ can be decomposed as
%%\begin{eqnarray}\label{eq:ucc_decomposition}
%%&&\exp(-i\theta\sigma_0^x\sigma_1^y)= \nonumber \\
%%&&R_0^x(-\frac{\pi}{2})R_1^{y}(\frac{\pi}{2})
%% CX_{1,0}R_0^z(\theta)CX_{1,0}R_0^x(\frac{\pi}{2})R_1^{y}(-\frac{\pi}{2})\nonumber \\
%%\end{eqnarray}
%%Here $CX_{i,j}$ represents CNOT gate that takes $i$-th qubit as control and $j$-th qubit as target. The corresponding variational quantum circuit is given in Fig.~\eqref{fig:H2_circuit_snake}.
For $\text{H}_2$, we consider an effective qubit Hamiltonian involves two qubits, following Ref.~\cite{hempel_18}. The unitary coupled cluster (UCC) ansatz is used, with unitary operator \[U(\theta)=\exp(-i\theta\sigma_0^x\sigma_1^y)\] performing on Hartree-Fock reference state is $\kets{01}$.
We chose $54$ points uniformly from bond lengths ranging from $0.25$ a.u. to $2.85$ a.u. Effective Hamiltonians corresponding to those bond lengths are obtained with OpenFermion~\cite{OpenFermion} Variational parameters are randomly initialized. We set $\alpha=0.1,\beta=3, \eta=0.5$ in the Eq.~\eqref{eq:collective_gd} (note that $\boldsymbol{A}$ depends on $\alpha$ and $\beta$).  Ground state energies at different bond lengths fit perfectly with ideal results. Remarkably, variational parameters for different bond lengths, while initialized randomly, quickly form a smooth curve and evolve to the target optimal values, as shown in Fig.~\eqref{fig:snake_algorithm}. This can be understood as a collective optimization process that exploits intricate relations between VQE tasks for  Hamiltonians with different bond lengths.

%\begin{figure}[h]
%	\includegraphics[width=1\columnwidth]{H2_circuit_snake.jpg}
%	\caption{Results of the snake algorithm for molecular hydrogen. (a) Variational quantum circuit for the unitary coupled cluster ansatz. (b).  Ground state energies at different lengths obtained from the snake algorithm. (c). Evolution of variational parameters in the optimization.}
%	\label{fig:H2_circuit_snake}
%\end{figure}

\subsection{Lithium hydride}
For LiH, STO-6G basis is used to construct the electronic Hamiltonian, which is mapped into a qubit Hamiltonian with BK transformation. Following Ref.~\cite{hempel_18}, three orbitals are chosen that the final qubit Hamiltonian evolves three qubits.
The UCC operator $U(\theta_1,\theta_2)=\exp(-i\theta_1\sigma_0^x\sigma_1^y)\exp(-i\theta_2\sigma_0^x\sigma_2^y)$
performs on an initial state $\kets{111}$. The operator can be taken as two UCCs, and each can be decomposed as in the Eq.\eqref{fig:one_two_circuit}.
%The resulting quantum circuit can be further simplified by combining sequential single-qubit rotations, and the final variational quantum circuit can be found in Fig.~\eqref{fig:LiH_circuit_snake}.
Effective Hamiltonians are calculated with OpenFermion from $50$ bond lengths, uniformly chosen from 0.3 a.u. to 5.0 a.u. Variational parameters are randomly initialized. We set $\alpha=0.1,\beta=3, \eta=0.2$ in the Eq.~\eqref{eq:collective_gd}. It can be seen in Fig.~\eqref{fig:snake_algorithm} that potential surface fit well with ideal results. Evolution of variational parameters turns to be rather impressive.  Unlike the case of molecular Hydrogen, there are two parameters for each VQE, and thus all points form a curve in the parameter space. The initial curve is random (due to random initialization) and is far away from the target. Nevertheless, the curve flows to the target curve by both shifting and changing its shape. Such a collective optimization process strikingly reminds of the behavior of a crawling snake.

\subsection{Helium hydride cation}
We now turn to consider Helium hydride cation, which is a more complicated molecular carrying  one positive charge. Under STO-3G basis, four qubits are required to describe the Hamiltonian~\cite{shen_17}. To capture essential quantum correlation, the UCC ansatz should include a two-particle scattering component~\cite{shen_17}. The unitary operator can be written as $
U(\theta_1,\theta_2,\theta_3)=\exp(-i\theta_3\sigma_0^x\sigma_1^x\sigma_2^x\sigma_3^y)\exp(-i\theta_2\sigma_1^x\sigma_3^y)\exp(-i\theta_1\sigma_0^x\sigma_2^y)$.
Effective Hamiltonians are calculated with OpenFermion from $30$ bond lengths, ranging from 0.25 a.u. to 2.5 a.u. Hyper parameters for the snake are set as $\alpha=0.1,\beta=3, \eta=0.2$ in the Eq.~\eqref{eq:collective_gd}. As there are three variational parameters, their evolution can be visualized as a crawling snake in a three dimensional space. Although initialized randomly, the snake becomes more smooth and moves to the target position. This again demonstrates the feature of the snake algorithm as a collective optimization process.

%\begin{figure}[h!]
%	\includegraphics[width=1\columnwidth]{LiH_circuit_snake.jpg}
%	\caption{Results of the snake algorithm for Lithium hydride. (a) Simplified variational quantum circuit for the unitary coupled cluster ansatz. (b). Ground state energies at different lengths (potential surface) calculated by the snake algorithm. (c). Evolution of variational parameters.}
%	\label{fig:LiH_circuit_snake}
%\end{figure}

\section{Nonconvex optimization of cVQE}
\label{sec:global_optimizer}
In the above, we have applied cVQE for solving ground-state energies of several molecules at different bond lengths. The process of optimization shows that parameters for different bond lengths evolve more smoothly, a remarkable feature of the snake algorithm for collective optimization. In this section, we further reveal that the snake algorithm trends for a global optimization, avoiding to be trapped at local minimums.

\begin{figure}[h!]
	\includegraphics[width=1\columnwidth]{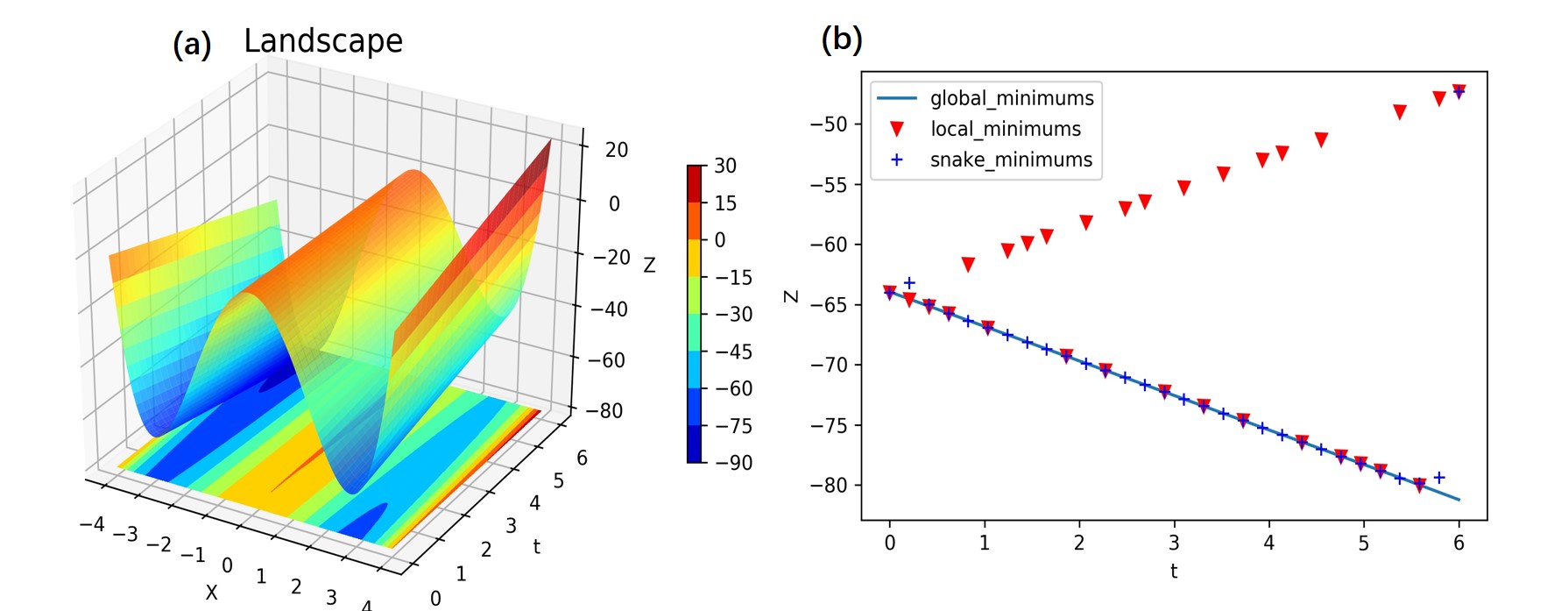}
	\caption{Optimization for a group of
		Styblinski-Tang functions parameterized with $t$ ($0\le t \le 6$).  $f(x;t)=\frac{1}{2}(x^4-16x^2+tx)$. (a). The landscape for a ST function has two minimums for fixed $t$.
		(b). Optimizations with gradient descent and the snake algorithm. It is shown that local minimums are often achieved by gradient descent why the snake algorithm can mostly achieve global minimums.}
	\label{fig:tang_1d}
\end{figure}

\begin{figure}[h!]
	\includegraphics[width=1\columnwidth]{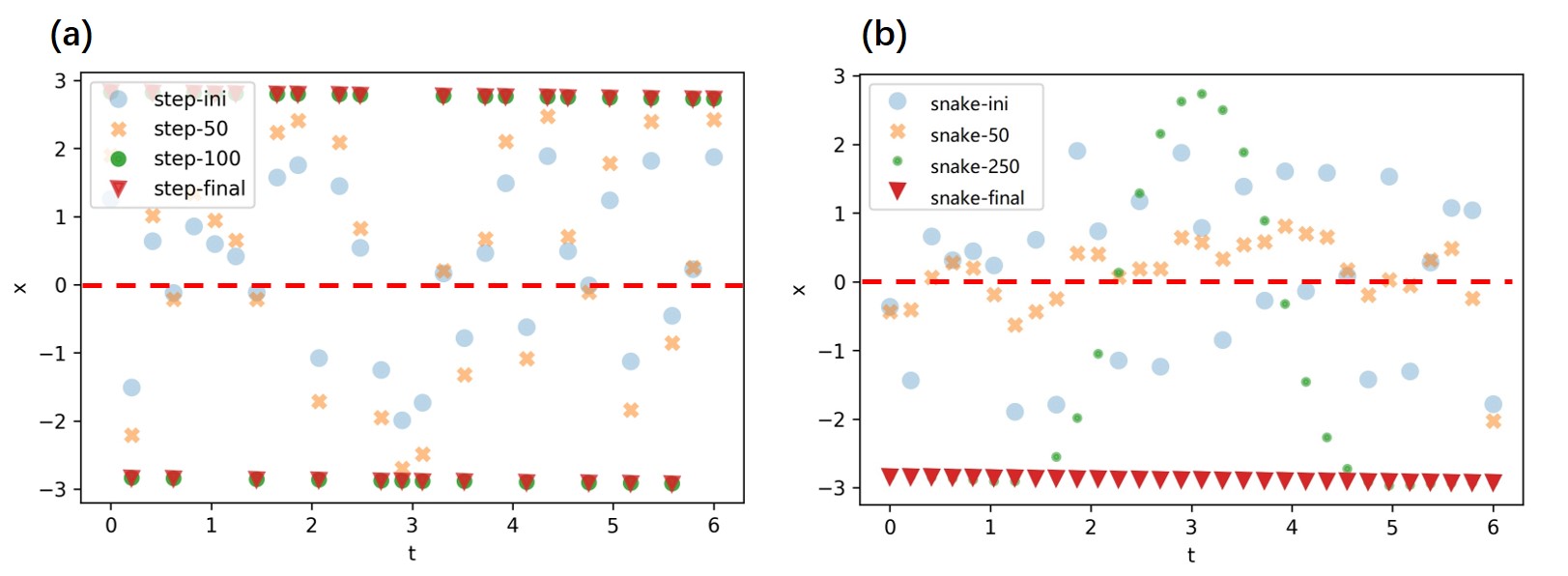}
	\caption{Optimization processes for nonconvex function. (a). Optimization by gradient descent. The flow of $x$ at fixed $t$ depends on the sign of initial value $x_0$, and if $x_0$ is positive then the optimization goes to the local minimum. (b). Optimization by the snake algorithm. It can be seen that almost all $x$ flow to global minimums collectively, even if they are initialized as positive and negative randomly. }
	\label{fig:tang_1d_snake}
\end{figure}

\begin{figure}[htbp] \centering
	\includegraphics[width=1.05\columnwidth]{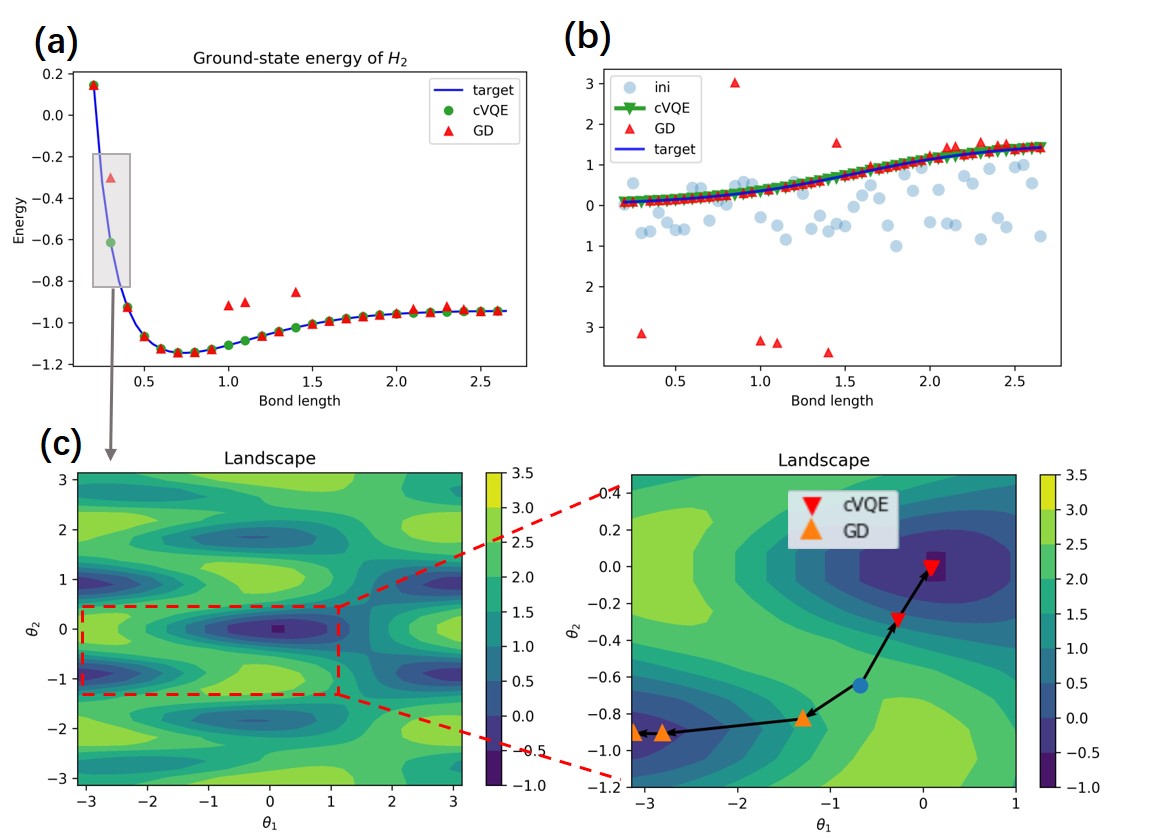}
	\caption{Nonconvex optimization for hydrogen molecule with the cVQE algorithm. (a). Optimization results using both the cVQE algorithm (marked as cVQE) and gradient descent~(marked as GD).(b). Evolving of parameters $\theta_1$ for the optimization process using the cVQE algorithm (greed line) and gradient descent(red triangular).(c). The landscape for VQE of hydrogen molecule has several different minimums. The random initial point flows to global minimum in cVQE metheod and to a local minimum in gradient descent.   }
	\label{fig:off_local_minimum}
\end{figure}

\subsection{Snake algorithm for nonconvex function}
We first use a toy example to illustrate how a collective optimization with the snake algorithm can avoid an optimization process to be trapped in local minimums.
We consider to minimize
the Styblinski-Tang (ST) function~\cite{styblinski_90}, a nonconvex function used to
benchmark optimization algorithms, defined as
$f(x)=\frac{1}{2}\sum_{i=1}^{N} x_i^4-16x_i^2+t_ix_i$.
To illustrate the mechanism of the snake algorithm for nonconvex optimization, we take $N=1$ and consider a group of ST functions, parameterized with $t$ as
$f(x;t)=\frac{1}{2}(x^4-16x^2+tx)$,
where $t\ge0$. For fixed $t$, there are two minimums
locating at $\pm x_0(t)$ and the global one locates at $-x_0(t)$ (assuming $x_0(t)>0$). However, those traps are deep that a optimizer may be easily trapped at local minimums, especially for optimizers based on gradient descents. The snake algorithm, although using gradient descent, can avoid this issue. As seen in Fig.~\eqref{fig:tang_1d}, most optimal points for different TS functions locate at global minimums. This is because all points are interconnected and can be  optimized collectively. Initially, there are some points located at traps of global minimums with random initialization. Then, those points will pull other point out of traps of local minimums, as seen in Fig.~\eqref{fig:tang_1d_snake}b. Such a mechanism can explain why the snake algorithm can be used as an optimizer for nonconvex function.

\subsection{Nonconvex optimization for VQE}
For variational quantum eigensolver, an expectation of Hamiltonian with regard to the variational wavefunction ansatz is in general a nonconvex function of variational parameters. As for illustration, we still consider the  hydrogen molecule with the same Hamiltonian as Eq.\eqref{ham:H2}, but the wavefunction ansatz is changed to
\[U(\theta_1,\theta_2)=\exp(-i\theta_2(a\sigma_0^x+b\sigma_1^x))\exp(-i\theta_1\sigma_0^x\sigma_1^y).\] Here $a$ and $b$ are fixed and we set $a=2,b=1.5$ for instance. Compared to the origin unitary coupled cluster ansatz, there is an extra term
$\exp(-i\theta_2(a\sigma_0^x+b\sigma_1^x))$. As seen in Fig.~\eqref{fig:off_local_minimum}b, the landscape has several different minimums. The global one locates at the center, corresponding to $\theta_2=0$. This is expected as the case of $\theta_2=0$ the wavefunction respects particle conservation, which is required for the system of hydrogen molecule.   A simple gradient descent as Eq.~\eqref{eq:gd} may lead to local minimums, once initially parameters of
$(\theta_1,\theta_2)$ are in traps of local minimums (Fig.~\eqref{fig:off_local_minimum}a). In fact, $\theta_2$ corresponding to those local minimums are far from zero, as seen in Fig.~\eqref{fig:off_local_minimum}c. However, the snake algorithm can perfectly overcome the issue of local minimums for optimizing VQE. During the optimization process, $\theta_1$ at different bond lengths evolve collectively. The curve connecting different $\theta_2$ becomes more smooth when approaching the target. Meanwhile, and all $\theta_1$ shrink to zero. Those present nice feature for nonconvex optimizations that are often met in VQE for quantum chemistry problems.

% \section{K-space analysis}
%Compared with Eq.~\ref{eq:gd} which updates only one point in the variational parameter space, Eq.~\ref{eq:collective_gd} is a collective gradient descent that updates $M$ points simultaneously.
%%The gradient flow can be seen as a snake crawls under external forces: nearby nodes of the snake move collectively.
%In fact, Eq.~\ref{eq:collective_gd} reduces to Eq.~\ref{eq:gd} for $\alpha=\beta=0$ (the step size is $\eta/\gamma$). It is useful to diagonalize $(\boldsymbol{A}+\gamma\boldsymbol{I})^{-1}$ and thus study those collective modes.
%\subsubsection{Periodic boundary condition}
%
%% change it into a k space. Different k components. and large k only contribute at first few stages.
%% write A down. give the largest and smallest number of A.
%But still those collective motions are coupled. How to make some approximations?
%
%Analytic very good! Largest eigenvalue=1.

\section{Discussion and summary}
\label{sec:summary}
Optimization is a key component for variational quantum eigensolvers. Here we have incorporated the snake algorithm for optimization of a group of VQE to find ground state energies for a molecular at different bond lengths. As the first step for collective optimization for quantum chemistry/many-body problems, it is expected that cVQE can be tested on more general wavefunction ansatzes. We have applied the unitary coupled cluster ansatz for quantum chemistry problem, and only consider a small number of variational parameters. For many quantum chemistry/many-body problems, more variational parameters are required and also other wavefunction ansatzes may be more suitable~\cite{liu_19}. The snake algorithm can be studied for such high dimensional optimization problem. It is expected that the snake algorithm can help to escape local minimums that often appear in a high-dimensional landscape.

%Secondly, it is interesting to make an in-detail connection between collective optimization and coupled differential equations, and then design more effective collective optimization methods from differential equations.  Recently, it has been revealed that advanced gradient descents in deep learning can be seen as a discrete version of second order differential equations~\cite{su_14}. Moreover, even more effective optimizer with speedup may be designed from high order differential equations.  For the snake algorithm, it can be viewed as a group of coupled differential equations. The question then is: how to design advanced snake algorithms from domain knowledge of partial differential equations?
%This is a very open problem and its investigations may help find more effective optimization methods for tasks using variational quantum circuits.

In summary, we have incorporated the snake algorithm to optimize variational quantum eigensolvers for a group of Hamiltonians. The cVQE has been used to solve ground states of molecules at different bond lengths simultaneously, which is enhanced by the collective optimization. Remarkably, we have demonstrated that the snake algorithm is a global optimizer, as the collective motion of variational parameters for different tasks can help pull parameters out of traps of local minimums.

% with many parameters, beyond unitary coupled cluster ansatz
% which can test the benefit of the snake algorithm

% a connection between partial differential equaitons and optimization
% especially on second order

% summary
\begin{acknowledgments}
	The authors thank the hosting by Peng Cheng Laboratory, where the manuscript is finalized. Thanks for Dr. Lei Wang's helpful discussion. This work is supported by the National Key Research and Development Program of China (Grant
	No. 2016YFA0301800), the National National Science Foundation of China (Grants No. 91636218, No.11474153, and No. U1801661), the Key Project of Science and  Technology of Guangzhou (Grant No. 201804020055).
\end{acknowledgments}

\appendix{
\section{Collective gradient descent}\label{app:colllective_gd}
In this section, we give details of deriving Eq.~\eqref{eq:collective_gd}.
The snake is determined from the least action principal. This is achieved by minimizing $L[\boldsymbol{\theta}(\lambda)]$. By Euler-Lagrange equation, this leads to a fourth-order differential equation,
\begin{equation}\label{eq:snake-diff}
\alpha\frac{\partial^2\boldsymbol{\theta}(\lambda)}{\partial^2\lambda}
+\beta\frac{\partial^4\boldsymbol{\theta}(\lambda)}{\partial^4\lambda}+\frac{dE}{d\boldsymbol{\theta}(\lambda)}=0.
\end{equation}
Here $E=\int_{\lambda_0}^{\lambda_T}\mathcal{E}(\boldsymbol{\theta}(\lambda))$. The last term of Eq.\eqref{eq:snake-diff} should be evaluated on a quantum processor, which makes Eq.~\eqref{eq:snake-diff} rather special and it is expected a solution with a hybrid quantum-classical algorithm.

The Eq.~\eqref{eq:snake-diff} should be solved numerally in a discrete version. $M$ different parameters are chosen uniformly from $[\lambda_0,\lambda_T]$ as $\{\lambda_1,\lambda_2,...,\lambda_M\}$, and $\lambda_T-\lambda_0=M\delta$.  Using finite difference, the second and forth orders of differentials turns to be Eq.~\eqref{eq:snake-diff},
\begin{eqnarray}
&&\left[\boldsymbol{\theta}(\lambda_{i+1})-2\boldsymbol{\theta}(\lambda_{i})+\boldsymbol{\theta}(\lambda_{i-1})\right]/{\delta^2}, \nonumber \\
&& \left[\boldsymbol{\theta}(\lambda_{i-2})-4\boldsymbol{\theta}(\lambda_{i-1})+6\boldsymbol{\theta}(\lambda_{i})-4\boldsymbol{\theta}(\lambda_{i+1})+\boldsymbol{\theta}(\lambda_{i+2})\right]/{\delta^4} \nonumber
\end{eqnarray}
respectively.
%we reach a Euler equation
%\begin{equation}\label{eq:matrix_pde}
%\boldsymbol{A}\boldsymbol{\theta}(\lambda_{i})+\frac{dE}{d\boldsymbol{\theta}(\lambda_{i})}=0,~~ i=1,2,...,M.
%\end{equation}=
Then we have
\begin{equation}\label{eq:matrix_pde}
\boldsymbol{A}\vec{r}_i+\frac{dE(\vec{r})}{d\vec{r}_i}=0,~~ i=1,2,...,N.
\end{equation}
For convenience we also introduce $\vec{r}_i=(\boldsymbol{\theta}_i(\lambda_1),\boldsymbol{\theta}_i(\lambda_2),...,\boldsymbol{\theta}_i(\lambda_M))$, and denote $E(\vec{r})=\sum_i\mathcal{E}(\boldsymbol{\theta}_{\lambda_i})$.
$\boldsymbol{A}$ is a pentadiagonal banded matrix with nonzero elements~(under the periodic condition), $\boldsymbol{A}_{i-2,i}=\boldsymbol{A}_{i,i-2}=\beta$,  $\boldsymbol{A}_{i-1,i}=\boldsymbol{A}_{i,i-1}=-\alpha-4\beta$,$\boldsymbol{A}_{i,i}=2\alpha+6\beta$, where $\delta^2$ and $\delta^4$ are absorbed accordingly.

% When $N=2$,  $(\vec{r}_1,\vec{r}_2)$ represents a parameterized curve on a plane.
Following Ref.\cite{kass_88}, the equation Eq.~\eqref{eq:matrix_pde} can be solved by introducing gradient flow (with an explicit Euler step, so it uses $\boldsymbol{A}\vec{r}_i^{t}$ instead of $\boldsymbol{A}\vec{r}_i^{t-1}$),
\begin{equation}\label{eq:matrix_gd}
-\frac{\vec{r}_i^{t}-\vec{r}_i^{t-1}}{\eta}=\boldsymbol{A}\vec{r}_i^{t}+\frac{dE(\vec{r}^{t-1})}{d\vec{r}_i},
\end{equation}
which leads to Eq.~\eqref{eq:collective_gd}.
%
%It should be noted that
%Eq.~\eqref{eq:collective_gd} is reduced to the gradient descent Eq.~\eqref{eq:gd} for a single variational quantum eigensolver
%by setting $\alpha=\beta=0$. In this sense, Eq.~\eqref{eq:collective_gd} generalizes gradient descent that optimizes all parameters for all tasks simultaneously. 	
	
\section{Hamiltonians and unitary cluster ansatz} \label{appdix:ham_openfermion}
Solving eigenvalues of electronic structures of molecules is the central problem for quantum chemistry. The ground-state energy is especially important as it largely determines the chemical properties of molecules.
The electronic Hamiltonian for a molecule consists of nuclear charges and electrons with Coulomb interactions. By Born-Oppenheimer approximation locations of nuclear are fixed. The electronic Hamiltonian is usually reformulated in the second quantized formulation, with a basis of $N$ molecular orbitals that are a linear combination of atomic orbitals. This can reduce the infinite dimension space of the original real space into a finite Hilbert space.  Solving eigenvalues and eigenstates can be done in this subspace. The dimensionality $N$ can be adjusted for the sake of precision demanded.

In the second quantization, the Hilbert space still grows exponentially with the number of orbitals $N$. It is important to only consider orbitals that contribute significantly to the low state energy. In practices, only active orbitals are considered, and inactive ones, such as occupied orbitals very close to the nuclear, or outside empty orbitals
are ignored. This leads to an effective electronic Hamiltonian that allows for feasible solutions.

The electronic Hamiltonian is fermionic and still can not be solved on a quantum processor. To map fermionic operators into qubit operators, one can refer to Jordan-Wigner transformation or Bravyi-Kitaev transformation. Those transformations are nonlocal and may introduce a tensor product of a string of Pauli matrices in the qubit Hamiltonian.

%\subsection{Effective qubit Hamiltonians}
We consider three kinds of molecules, molecular hydrogen and Lithium hydride, and helium hydride cation. Their qubit Hamiltonians with varying bond lengths are calculated with the open source software OpenFermion~\cite{OpenFermion}, following setups in Ref.\cite{hempel_18} for hydrogen and Lithium hydride, and Ref.\cite{shen_17} for helium hydride cation.

For $\text{H}_2$, and STO-3G minimal basis are adopted, the final effective qubit Hamiltonian involves two qubits, which can be written as
\begin{eqnarray}\label{ham:H2}
H_{\text{H}_2}(\lambda)&=&c_0(\lambda)\mathcal{I}+c_1(\lambda)\sigma_0^z+c_2(\lambda)\sigma_1^z +c_3(\lambda)\sigma_0^z\sigma_1^z \nonumber \\
&+&c_4(\lambda)\sigma_0^x\sigma_1^x+c_5(\lambda)\sigma_0^y\sigma_1^y.
\end{eqnarray}
Here, coefficients $c_i(\lambda)$ depend on the bond length $\lambda$ and their values can be found in the code. The Hartree-Fock reference state is $\kets{01}$.

For LiH, STO-6G basis is used to construct the electronic Hamiltonian, which is mapped into a qubit Hamiltonian with BK transformation. Following ref.cite, three orbitals are chosen that the final qubit Hamiltonian evolves three qubits,
\begin{eqnarray} \label{ham:LiH}
&&H_{\text{LiH}}(\lambda) \nonumber \\
&&=c_0(\lambda)\mathcal{I}+c_1(\lambda)\sigma_0^z+c_2(\lambda)\sigma_1^z +c_3(\lambda)\sigma_2^z +c_4(\lambda)\sigma_0^z\sigma_1^z\nonumber \\ &&+c_5(\lambda)\sigma_0^z\sigma_2^z
+c_6(\lambda)\sigma_1^z\sigma_2^z+c_7(\lambda)\sigma_0^x\sigma_1^x+c_8(\lambda)\sigma_0^x\sigma_2^x \nonumber \\
&&+c_9(\lambda)\sigma_1^x\sigma_2^x
+c_10(\lambda)\sigma_0^y\sigma_1^y+c_{10}(\lambda)\sigma_0^y\sigma_2^y+c_{11}(\lambda)\sigma_1^y\sigma_2^y \nonumber \\
\end{eqnarray}
The reference state is $\kets{001}$.

For the above two effective qubit Hamiltonians, we adopt simple unitary coupled cluster ansatz~\cite{chan_04,taube_06,yung_14,shen_17,hempel_18}, which can establish entanglement between different qubits and thus take quantum correlations into account.
For $H_2$, the unitary operator is \[U(\theta)=\exp(-i\theta\sigma_0^x\sigma_1^y)\] and the wavefunciton ansatz is $U(\theta)\kets{01}$. The parameter $\theta$ can character the degree of entanglement the electron and the hole. For LiH, the UCC ansatz is \[U(\theta_1,\theta_2)=\exp(-i\theta_2\sigma_0^x\sigma_2^y)\exp(-i\theta_1\sigma_0^x\sigma_1^y),\]
and the wavefunciton ansatz is $U(\theta_1,\theta_2)\kets{111}$.
$U(\theta_1,\theta_2)$ can be decoupled as two entanglers that establish an entanglement of the zeroth and the first orbitals, the zeroth and the second orbitals, respectively.
Two parameters $\theta_1$ and $\theta_2$ characterize the degrees
of entanglement, correspondingly.

We also consider helium hydride cation ($\text{HeH}^+$). Under STO-3G basis, its qubit Hamiltonian includes both two, three and four spin interactions.
%\begin{eqnarray}
%&&H_{\text{HeH}^+}(\lambda)=c_0I + \sum_{i=0}^{3}c_iZ_i+\sum_{i,j=0}^{3} (z_{ij}Z_iZ_j+\nonumber \\
%&&x_{ij}X_iX_j+y_{ij}Y_iY_j )
%+\sum_{i=0}^{1}t_3(X_iZ_{i+1}X_{i+2}+Y_iZ_{i+1}Y_{i+2})\nonumber \\
%&&+f_0X_0X_1Y_2Y_3+f_1Y_0Y_1X_2X_3+f_2X_0Y_1Y_2X_3\nonumber \\
%&&+f_3Y_0X_1X_2Y_3
%+f_4X_0Z_1X_2Z_3
%+f_5Z_0X_1Z_2X_3\nonumber \\
%&&+f_6Y_0Z_1Y_2Z_3+f_7Z_0Y_1Z_2Y_3. \nonumber \\
%\end{eqnarray}

\begin{widetext}
\begin{eqnarray}
%\begin{split}
H_{\text{HeH}^+}(\lambda)&=&c_0I + \sum_{i=0}^{3}c_iZ_i+\sum_{i,j=0}^{3} (z_{ij}Z_iZ_j+x_{ij}X_iX_j+y_{ij}Y_iY_j )+\sum_{i=0}^{1}t_3(X_iZ_{i+1}X_{i+2}+Y_iZ_{i+1}Y_{i+2})\nonumber \\
&&+f_0X_0X_1Y_2Y_3+f_1Y_0Y_1X_2X_3+f_2X_0Y_1Y_2X_3
+f_3Y_0X_1X_2Y_3\nonumber \\
&&+f_4X_0Z_1X_2Z_3
+f_5Z_0X_1Z_2X_3+f_6Y_0Z_1Y_2Z_3+f_7Z_0Y_1Z_2Y_3. \nonumber \\
%\end{split}
\end{eqnarray}
\end{widetext}
An UCC ansatz for $H_{\text{HeH}^+}(\lambda)$ should consider both first and second excitation. Following ref.~\cite{shen_17}, we use
\begin{eqnarray}
&&U(\theta_1,\theta_2,\theta_3)= \nonumber\\
&&\exp(-i\theta_3\sigma_0^x\sigma_1^x\sigma_2^x\sigma_3^y)\exp(-i\theta_2\sigma_1^x\sigma_3^y)
\exp(-i\theta_1\sigma_0^x\sigma_2^y).\nonumber\\
\end{eqnarray}
The wavefunciton ansatz is $U(\theta_1,\theta_2,\theta_3)\kets{0011}$.

To implement the above ansatzes on quantum processors, we need to decompose Hamiltonian evolution of one-particle transition and two-particle transition into a set of universal quantum gates involving single-qubit rotations and two-qubit CNOT gate, as can be seen in Fig~\eqref{fig:one_two_circuit}
\begin{figure}[h!]
	\includegraphics[width=1\columnwidth]{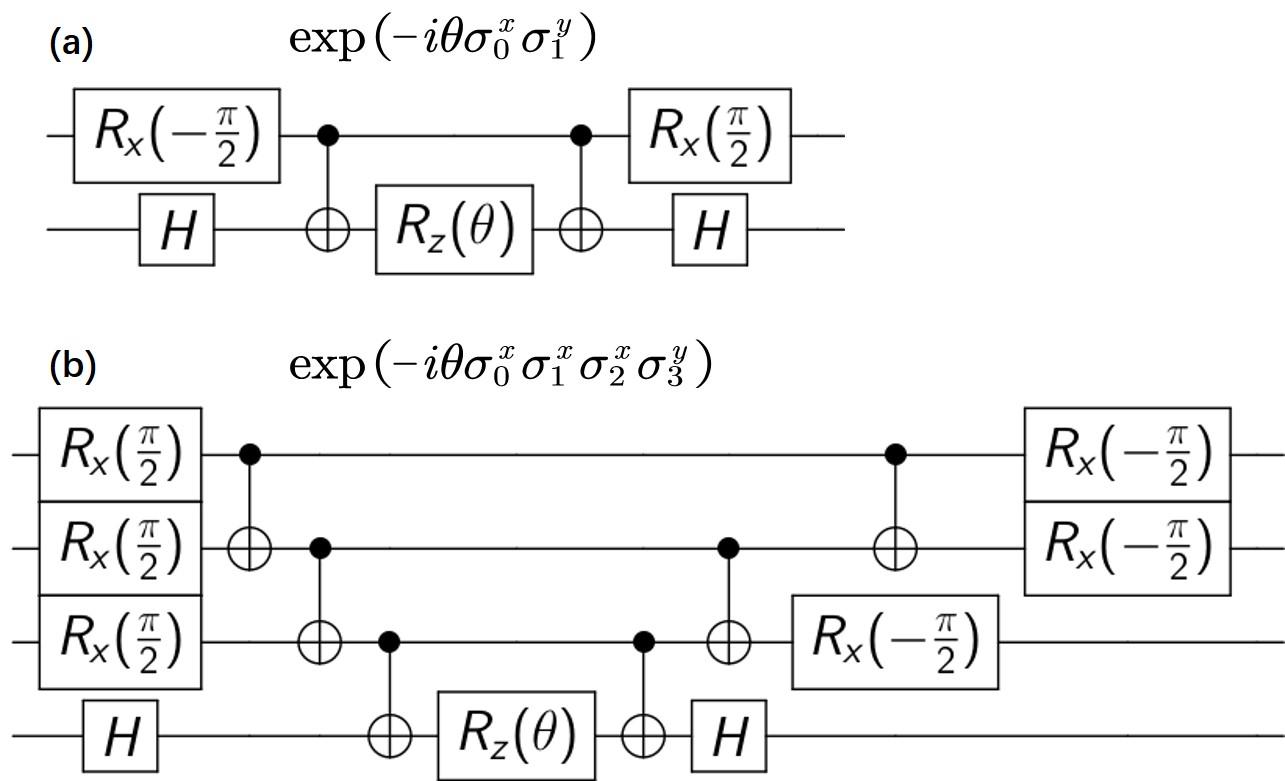}
	\caption{Decomposition of basic operator in unitary coupled cluster ansatz into a set of universal quantum gates.}
	\label{fig:one_two_circuit}
\end{figure}
The decomposition makes the UCC operator implementable on quantum chips. Moreover, variational parameters only appear in a single-qubit rotation $R_z(\theta)$. Thus, an analytic gradient can be evaluated using the shift rule.	
}

\end{document}